\RequirePackage{fix-cm}
\documentclass[twocolumn]{svjour3}  
\smartqed

\usepackage{amssymb}
\usepackage{bm}
\usepackage{graphicx}
\usepackage{caption}
\usepackage{subcaption}
\usepackage{bigints}
\usepackage{amssymb}

\newcommand{\beq}{\begin{equation}}
\newcommand{\eeq}{\end{equation}}
\newcommand{\ba}{\begin{array}}
\newcommand{\ea}{\end{array}}
\newcommand{\bqa}{\begin{eqnarray}}
\newcommand{\eqa}{\end{eqnarray}}
\newcommand{\bea}{\begin{eqnarray}}
\newcommand{\eea}{\end{eqnarray}}

\begin{document}
\title{Analysis and comparative study of non-holonomic and quasi-integrable deformations of the Nonlinear Schr\"odinger Equation}
\author{Kumar Abhinav \and Partha Guha \and Indranil Mukherjee}

\institute{K. Abhinav (Corresponding Author)\at
              The Institute for Fundamental Study (IF), Naresuan University, Phitsanulok 65000, Thailand \\
              Tel.: +66-5596-8730\\
              \email{kumara@nu.ac.th}           
           \and
           P. Guha \at
              S. N. Bose National Centre for Basic Sciences, JD Block, Sector III, Salt Lake, Kolkata - 700098,  India\\
              \email{partha@bose.res.in}
              \and
        I. Mukherjee \at
        Department of Natural Science,M A K Azad University of Technology, West Bengal, Haringhata, Nadia, India\\
        \email{indranil.m11@gmail.com}
}


\maketitle
\begin{abstract}
The non-holonomic deformation of the
nonlinear Schr\"odinger equation, uniquely obtained from
both the Lax pair and Kupershmidt's bi-Hamiltonian [Phys. Lett. A {\bf 372}, 2634 (2008)] approaches, is compared with the quasi-integrable deformation of the same system [Ferreira {\it et. al.} JHEP {\bf 2012}, 103 (2012)]. It is found that these two deformations can locally coincide only when the phase of the corresponding solution is discontinuous in space, following a definite phase-modulus coupling of the non-holonomic inhomogeneity function. These two deformations are further found to be not gauge-equivalent in general, following the Lax formalism of the nonlinear Schr\"odinger equation. However, asymptotically they converge for localized solutions as expected. Similar conditional correspondence of non-holonomic deformation with a non-integrable deformation, namely, due to local scaling of the amplitude of the nonlinear Schr\"odinger equation is further obtained.
\keywords{Nonlinear Schr\"odinger equation \and Quasi-integrable deformation \and Non-holonomic deformation \and Solitons}
\end{abstract}

\section{Introduction}
Integrable partial differential equations appearing in field theory are best studied via the Lax pair method (zero curvature condition). They are deemed integrable if they contain
infinitely many conserved quantities responsible for the stability of the soliton solutions \cite{Das}.
In particular, these constants of motion uniquely define the dynamics
of the system, rendering the corresponding equations to be completely solvable.
The Nonlinear Schr\"odinger (NLS) equation, in one space and
one time (1+1) dimensions, is one such system that further incorporates semi-classical soliton solutions which are physically realizable. These solitons have high degree of symmetry that mandates
infinitely many conserved quantities. Such stable solutions of integrable
models are subjected to the zero-curvature condition \cite{Z1,Z2} involving connections that constitute the Lax pair which linearize the non-linear system.

\paragraph*{}However, real physical systems do not posses infinite
degrees of freedom and thus a corresponding field-theoretical model cannot be integrable
in principle. On the other hand such systems do possess solitonic
states very similar to those of integrable models, {\it e. g.}, sine-Gordon
\cite{SSG0}. Therefore the study of continuous physical systems as slightly
deformed integrable models makes conceptual sense. Recently, it was shown that
the sine-Gordon model can be deformed and following suitable approximation
leads to a finite number of conserved quantities \cite{f1}. However, a class
of deformed defocussing NLS \cite{New1} and SG \cite{New01} models display
an infinite subset of the charges to be conserved, leaving out an infinite
number of anomalous charges. In these systems, an almost flat connection
induces anomaly in the zero-curvature condition, rendering them to be {\it
quasi-integrable}. Exact dark \cite{New1} and bright \cite{New2} soliton
configurations of the quasi-NLS system have very recently been obtained, the latter
having infinite towers of {\it exactly} conserved charges, making it very
close to be integrable. Ferreira {\it et al} \cite{1} considered
modifications of the NLSE to investigate the concept of quasi-integrability,
where they perturbed the NLS potential (non-linearity) as $V ~ (|\psi|^2)^{2+\varepsilon}$ by the parameter $\varepsilon$ to show that such
models possess an infinite number of charges which are conserved only
asymptotically.

\paragraph*{}On the other hand, different systems which are completely integrable have been found to be related through certain deformations known as non-holonomic deformations (NHD). As a concrete example of this class of deformation, we momentarily digress to the work of Karasu-Kalkani {\it et al} \cite{KK} who demonstrated that the integrable 6th order KdV equation represented a NHD of the celebrated KdV equation. The equation is given by,
\beq
(\partial_x^3+8u_x \partial_x+4u_{xx})(u_t+u_{xxx}+6u_x^2)=0.\label{E1}
\eeq
With the change of variables $v=u_x$ and $w=u_t+u_{xxx}+6u_x^2$ the above equation can be rewritten as a pair:
\bqa
&&v_t+v_{xxx}+12vv_x-w_x=0\nonumber\\
{\rm and}\quad && w_{xxx}+8vw_x+4wv_x=0.\label{E2}
\eqa
The authors of \cite{KK} obtained the Lax pair as well as an auto-B\"acklund transformation corresponding to Eq.s \ref{E2} and claimed that these equations were different from a KdV equations having self consistent sources. They explored corresponding higher-order symmetries, conserved densities and Hamiltonian formalism. It is a recurring characteristic that the integrability properties of two systems can be completely different even if they are connected through NHD.

\paragraph*{}The terminology ``non-holonomic deformation" was introduced by Kuperschmidt \cite{Kuppershmidt}, who re-scaled $v$ and $t$ to modify Eq.s \ref{E2} as:
\bqa
&&u_t-6uu_x-u_{xxx}+w_x=0\nonumber\\
{\rm and}\quad && w_{xxx}+4uw_x+2u_xw=0.
\eqa
The above pair of equations can now converted into a bi-Hamiltonian system,
\bqa
&&u_t=B_1 \left( \frac{\delta H_{n+1}}{\delta u}  \right)  - B_1(w) \nonumber\\
&&\quad=B_2 \left( \frac{\delta H_n}{\delta u} \right)  - B_1(w)\nonumber\\
{\rm and}\quad &&B_2(w)=0,\label{E4}
\eqa
with the Hamiltonian operators,
\beq
B_1\equiv \partial \equiv \partial_x   \quad{\rm and}\quad
B_2\equiv \partial^3+2(u\partial + \partial u),
\eeq
where $H_n$s denote the conserved densities. First few of the conserved densities for the KdV case are given by $H_1$ = u, $H_2 = \frac{1}{2}u^2$, $H_3 = u^3 - \frac{1}{2}u_x^2$ etc. and $w$ represents the deformation to obtain the 6th order KdV system. Following Ref. \cite{KK}, the formalism in Eq.s \ref{E4} allows for NHD of {\it any} bi-Hamilotonian system. This understanding is carried forward to the NLS system in particular which models many well-known physical systems. However, we will further adopt a more convenient approach to obtain the corresponding NHD in the following, though it is to be noted that Kuperschmidt's formulation depicted above sets up the generic integrability structure of this class of deformation.

\paragraph*{}The NHD for field theoretical integrable models is known to be characterized by constraints in the form of nonlinear differential equation(s) involving {\it only} $x$-derivatives of the perturbing function(s), obtained by deforming the original integrable equation \cite{3}. This type of integrable deformation is relatively new, which also allows to build infinite dimensional framework of Euler-Poincar\'e-Suslov theory \cite{GuhaJPA,GuhaRMP}. It is well-known that actual physical systems can directly or indirectly be represented by nonlinear equations. For example, the NLS equation itself is the mean-field description of a four-Fermi interaction \cite{FW} responsible for phenomena like superconductivity, superfluidity and Bose-Einstein condensation. Further, the same equation is the continuum representation of the one dimensional Heisenberg XXX spin system \cite{Laxmanan} through the Hasimoto map \cite{Hasimoto}. Such maps further links more generalized systems like one dimensional inhomogeneous Heisenberg XXX spin system to an integro-differential NLS-type equation \cite{Balakrishnan1} which is integrable \cite{Balakrishnan2} and can be identified as an NHD of the standard NLS equation \cite{OwnEPJB}. Further a different NHD of the NLS system can be mapped \cite{OwnEPJB} to quantum vortex filament moving with a drag \cite{Shiva}. It is also known that through Bethe ansatz the one dimensional Heisenberg XXZ spin system is related to the KdV equation \cite{XXZKdV}. These are compelling motivations towards the study of non-holonomic nonlinear equations, especially of the NLS type, as it is highly likely to correlate otherwise completely independent physical systems and thereby to improve their understanding.

\subsection{The purpose and structure of the paper}
The purpose of the present work is to seek a connection between two different types of deformations explained above, non-holonomic  and quasi-integrable, subjected to the NLS system. This is of interest owing to the fact that quasi-deformed systems retain integrability in the asymptotic limit, whereas NHD maintains integrability absolutely, albeit constraint at {\it higher} order. Even {\it locally}, the quasi-integrable deformation (QID) could appear as local inhomogeneity of the NLS system, of the same form as in NHD. Therefore, though essentially different, these two deformations can be related under certain conditions. In doing so, further, a clear cut demonstration of the distinction between the two deformations is expected. It is found that they indeed can be identified, both locally as the phase of the quasi-deformed solution becomes discontinuous in space,
and asymptotically where moduli of the solutions in both cases as well as that of the deformation inhomogeneity become constants as the quasi-deformed system regains integrability. Locally, in general, these two deformations mandates distinct analytic and physical nature with a clear criteria as only one of them is integrable. In addition the equivalence between the Lax pair and the Kupershmidt ansatz approaches for non-holonomic NLS system is further explicated in the process. 

\paragraph*{}The rest of the paper is organized as follows: Section 2 demonstrates the NHD of the NLS system; by using both the Lax pair approach and the Kupershmidt's bi-Hamiltonian prescription and some further discussions in its successive subsections. Section 3
describes a few types of QIDs of the same system in the defocussing case. Section 4 deals with derivation of explicit condition(s) for correspondence between QID of the difocussing NLS system with its NHD. We further demonstrate in detail the extent of their general incompatibility.
This conditional compatibility of NHD is further seen to extend to local scaling of
the NLS amplitude. We conclude in Section 5 by pointing out possible outcomes.

\section{NHD of the NLS equation}\label{S2}
It would be pertinent to explain what exactly is meant by the Non-Holonomic Deformation of integrable systems. Perturbation generally disturbs the integrability of a system. However, when we consider NHD of an integrable system, the perturbation is such that under suitable differential constraints on the perturbing function, the system maintains its integrability. The constraints are furnished in the form of differential relations which are non-holonomic in nature.

\paragraph*{}To impose a NHD, one starts with the concerned Lax pair of the system, keeping the space part $U(\lambda)$ unchanged but modifying the temporal component $V(\lambda)$, $\lambda$ being the spectral parameter. This implies that the scattering problem remains unchanged, but the time evolution of the spectral data becomes different in the perturbed models. To retain integrability the non-holonomic constraints have to be affine in velocities prohibiting explicit velocity dependence after the deformation. This insists on the deformation being exclusive to the temporal Lax component as it is not acted upon by a time derivative to yield the dynamical equation \cite{NHD1}. The explication of this particular point will be reflected in the next section by the exclusive time-dependence of the parameters when NHD can be identified with QID.
Corresponding to these deformed systems, it is possible to generate some kind of two-fold integrable hierarchy. One method is to keep the perturbed equations the same but increase the order of the differential constraints in a recursive manner, thus generating a new integrable hierarchy for the deformed system. Alternatively, the constraint itself may be kept fixed at its lowest level, but the order of the original equation may be increased in the usual way, thereby leading to new hierarchies of integrable systems.  Indeed we will demonstrate in the next sections that one can derive NH NLS equation in two equivalent methods, using Lax pair representation and the bi-Hamiltonian method respectively.

\paragraph*{}The method of adding extra terms in the Lax pair (in this case, the temporal component of the Lax pair) has been adopted in some earlier cases also \cite{GS,FGGS} wherein some integrable generalizations of the Toda system generated by flat connection forms taking values in higher Z-grading subspaces of a simple Lie algebra was considered. However, in case of the non-holonomic deformation, the additional relations generated due to inclusion of extra terms in the temporal Lax component are treated as differential constraints imposed on the deformed systems and not as equations involving new field variables. 

\subsection{Non-holonomic deformation using Lax method}
The spatial and temporal components of the Lax pair for the NLS equation are respectively given by,

\bqa
&&U=i \lambda \sigma_3 +q \sigma_+ + r \sigma_-,\nonumber\\
&&V_O=-i\left(\lambda^2+\frac{1}{2}qr\right)\sigma_3 + \lambda(q \sigma_+ + r\sigma_-) \nonumber\\
&&\quad+ \frac{i}{2}\left(q_x\sigma_+ -r_x\sigma_-\right). \label{N01}
\eqa
Then the paired NLS equation, in terms of both the dynamical variables $q$ and $r$ take the forms:
\beq
q_t -\frac{i}{2}q_{xx}+iq^2r=0,\quad{\rm and}\quad r_t +\frac{i}{2}r_{xx}-ir^2q=0,
\eeq
which can be obtained as consistency equations, by imposing the usual zero-curvature condition:
\beq
U_t-V_{O,~x}+[U,V_O]=0. \label{N02}
\eeq
The only scale present in the system is the spectral parameter $\lambda$, defining the corresponding solution space. In order to deform the temporal component $V_O$ so that integrability is preserved through the flatness condition (Eq. \ref{N02}), it is intuitively
obvious that the deformation part has to be a function of $\lambda$. We consider the following additive
deformation term:

\bqa\label{N102}
&&V_D=\frac{i}{2}\lambda^{-1} G^{(1)},\nonumber\\
{\rm where},\quad &&G^{(1)}=a\sigma_3+g_1\sigma_++g_2\sigma_-,
\eqa
so that the resultant overall temporal Lax component takes the form,

\beq   
\tilde{V}=V_O + V_D,
\eeq
thereby changing zero curvature condition to,
\beq
F_{tx}=U_t-\tilde{V}_x+[U,\tilde{V}]=0,\label{ZC}
\eeq
in order to keep the system integrable.
The adopted deformation of Eq. \ref{N102} contains only ${\cal O}\left(\lambda^{-1}\right)$ terms. Deformation terms of ${\cal O}\left(\lambda^{n\geq 0}\right)$
only leads to additional perturbed dynamical systems at each order, and
vanish when the terms are substituted order-by-order. This is because the NLS equations arise from contributions at
${\cal O}\left(\lambda^{0}\right)$ in the flatness condition, and the presence of any higher order contribution is
decoupled from the corresponding dynamics. Therefore the higher order deformations end-up yielding trivial identities that eventually eliminates all the NLS contributions at ${\cal O}\left(\lambda^{n\geq 0}\right)$ which can easily be verified. Hence the deformation of Eq. \ref{N102} can be considered as a general one.

\paragraph*{}The ${\cal O}\left(\lambda^{0}\right)$ terms in the zero-curvature condition leads to the {\it deformed} pair of the NLS equations:

\bqa\label{N03}
&&q_t-\frac{i}{2}q_{xx}+iq^2r=-g_1 \nonumber\\
{\rm and}\quad &&r_t+\frac{i}{2}r_{xx}-iqr^2=g_2.
\eqa
As expected trivially, inhomogeneous terms $g_{1,2}$ are introduced in the NLS system. Such equations are already known to be integrable, and thus, the primary objective is achieved.

\paragraph*{}The ${\cal O}\left(\lambda^{-1}\right)$ sector in the flatness condition, on equating the coefficients of the generators $\sigma_3$, $\sigma_+$ and $\sigma_-$, yields the constraints:

\bqa\label{N04}
&&a_x - q g_2 +r g_1 = 0, \quad g_{1x} + 2aq = 0\nonumber\\
{\rm and}\quad&& g_{2x} - 2ar = 0,
\eqa
on the functions $a$, $g_1$ and $g_2$ respectively.
The last two of the above set of three equations can be combined to yield the expression $$ rg_{1x} + qg_{2x} = 0.$$
Again, all these equations can be combined, through mutual substitution, to give
rise to the {\it differential constraint}:
\beq
\hat{L}(g_1, g_2) = r g_{1xx} + q_x g_{2x} + 2qr (q g_2 - r g_1) = 0,\label{N05}
\eeq
which can be used to eliminate the deforming functions $g_1$ and $g_2$ in Eqs. \ref{N03}, to obtain
a new higher order equation as,
\bqa
&&r (q_t - \frac{i}{2} q_{xx}+ i q^2 r)_{xx} = q_x (r_t + \frac{i}{2} r_{xx} - i q r^2)_x \nonumber\\
&&\qquad\qquad\qquad\qquad\qquad+  2qr \Big[q (r_t + \frac{i}{2} r_{xx} - i q r^2)\nonumber\\
&&\qquad\qquad\qquad\qquad\qquad + r(q_t - \frac{i}{2} q_{xx} + iq^2 r)\Big]. \label{N06}
\eqa
This equation is subjected to the dynamics of Eq.s \ref{N03}, and therefore do not yield any new dynamics, and eventually
reflects only the constraint in a different form. This is in accord with the previous argument that no term, with power of
$\lambda$ other than that responsible for yielding Eqs. \ref{N03}, can yield dynamics of the NLS system, as it will violate
the overall integrability of the system itself.

\paragraph*{}The constraint of Eq. \ref{N05}, characterizing the deformation, is {\it non-holonomic} in nature as it contains differentials of corresponding variables. Noticeably such a constraint solely arises from the terms with negative power of the spectral parameter. Further, explicit forms of the local functions $a$,
$g_1$ and $g_2$ are not necessary to establish the integrability, and they represent a class that satisfies the constraint in Eq. \ref{N05}.
In other words, constraints arise form ${\cal O}\left(\lambda^{-1}\right)$ contributions, and {\it additionally}
restrict the allowed values of $q$ and $r$ of the deformed dynamics at ${\cal O}\left(\lambda^0\right)$. The former is
necessary for integrability while the latter ensures that any solution of Eq.s \ref{N03} is valid for Eq. \ref{N06}. For the
defocussing case $r=q^*$, which is of main interest of present comparative study with QID, Eq.s
\ref{N04}, \ref{N05} and \ref{N06} exactly reduces to those found for NH NLS system in Ref. \cite{Kundu01} where the
exact solution was constructed in a phenomenological way. Therein a further demonstration of {\it twofold integrable hierarchy} using the
constraint Eq.s \ref{N04} in defocussing case was performed, including a higher order NLS equation with constrained
perturbation. This is discussed in a general framework considering ${\cal O}\left(\lambda^{-1}\right)$ perturbation
below. The other hierarchy appears as the tri-Hamiltonian formulation of self-induced transparency equations
\cite{Fordy} with explicit integrability and isospectral flows and thereby establishing the same for the NLS counterpart.
This is rather a special case of NHD where the flows regarding deformed equation and constraints do commute, unlike the
general scenario where they do not, as demonstrated for the KdV6 system \cite{Kuppershmidt}. The non-holonomic deformation of DNLS equation for controlling the optical soliton in doped fibre media is discussed in \cite{Kundu02} while the role of self-induced transparency of solitons in erbium-doped fiber waveguide is dealt with in \cite{NKKS}. The coexistence of two different types of solitons, {\it viz.} the self-induced transparency soliton and the non-linear Schrodinger one is examined in \cite{NYK}.

\subsection{Non-holonomic deformation via bi-Hamiltonian method}
We now apply the ansatz due to Kupershmidt \cite{Kuppershmidt} to derive the non-holonomic deformed NLS equations as well as the constraints on the deforming variables themselves. The NLS equations are written, after slight rescaling of the variables to resemble Kupershmidt's definitions, as:

\bqa
&& q_t = q_{xx}- 2q^2r,\nonumber\\
{\rm and}\quad && r_t = -r_{xx}+ 2r^2q.\nonumber
\eqa
The bi-Hamiltonian structures of the pair of NLS equations are given by:

\bqa
			&& B^1=  \left(\begin{array}{cc}
0 & -1 \\
                                           1  & 0  \\
                                    \end{array}\right),\nonumber\\
&& B^2 = \left(\begin{array}{cc}
2q\partial_x^{-1}q & \partial_x - 2q\partial_x^{-1}r \\
                                            \partial_x - 2r\partial_x^{-1}q &  2r\partial_x^{-1}r \\
                                    \end{array}\right),
\eqa
and the corresponding conserved densities are:
\beq
H^1 = q_xr_x+ q^2r^2 \quad{\rm and}\quad H^2 = q_xr.
\eeq

\paragraph*{}Introducing $w_1$ and $w_2$ as the deforming variables and following the Kupershmidt ansatz, the pair of NLS equations under non-holonomic deformation can be written in the following way:

\bqa
\left(
\begin{array}{cc}
q\\
r
\end{array}
\right)_t&=&B^1\left(
\begin{array}{cc}
\frac{\delta}{\delta q}\\
\frac{\delta}{\delta r}
\end{array}
\right)H^1-B^1\left(
\begin{array}{cc}
w_1\\
w_2
\end{array}
\right)\nonumber\\
&=&B^2\left(
\begin{array}{cc}
\frac{\delta}{\delta q}\\
\frac{\delta}{\delta r}
\end{array}
\right)H^2-B^1\left(
\begin{array}{cc}
w_1\\
w_2
\end{array}
\right),
\eqa
leading to the final forms,

\beq
q_t=  q_{xx} - 2q^2 r +w_2  \quad{\rm and}\quad r_t=  -r_{xx} + 2qr^2 - w_1.
\eeq
The constraint conditions on the deforming variables $w_{1,2}$, which can easily be identified with $g_{1,2}$ of the previous subsection, are obtained in the integro-differential form by setting,

\beq
B^2\left(
\begin{array}{cc}
w_1\\
w_2
\end{array}
\right)=0,
\eeq
that leads to the conditions:

\bqa
&&w_{1x} +  2r \partial_x^{-1}(rw_2 - qw_1) = 0\nonumber\\
{\rm and}\quad && w_{2x} +  2q \partial_x^{-1}(qw_1 - rw_2) = 0.
\eqa
On multiplying first of the above equations by $q$ and the second by $r$ and adding together we obtain,

\beq
qw_{1x} + rw_{2x} = 0.
\eeq
This is exactly similar to the relation obtained among the field variables and deforming variables obtained using the Lax pair method. This is probably the first time that the equivalence between the Lax pair and the bi-Hamiltonian method for the NHD of an integrable systems is explicitly obtained.

\subsection{Further consideration of the Non-holonomic deformation}
Although we have considered only an ${\cal O}\left(\lambda^{-1}\right)$ deformation of the NLS system, this process can
be extended to ones with higher negative power of $\lambda$ as a hierarchical deformation structure \cite{Kundu01}, with
additional constrained dynamics. To show that, we next consider a `higher' order deformation by taking,

\beq
V_D(\lambda) = \frac{i}{2}\left (\lambda^{-1} G^{(1)} + \lambda^{-2} G^{(2)}\right), \label{N07}
\eeq
where the second order contribution $G^{(2)}$ has the form:

\beq
G^{(2)} = b\sigma_3 + f_1 \sigma_+ + f_2 \sigma_-. \label{N08}
\eeq
As a consequence the zero-curvature condition, with the new $V_D$, leads to the following results:

\begin{enumerate}
\item No change occurs in the deformed NLS equations, as the new contribution $V_D$ is of order ${\cal O}\left(\lambda^{-2}\right)$.\\
\item Picking up the terms in $\lambda^{-1}$ and equating the coefficients of the generators $\sigma_3$, $\sigma_+$ and $\sigma_-$ successively,
we are led to the following new set of conditions:

\bqa
a_x - q g_2 + r g_1 &=&0,\nonumber\\
g_{1x} + 2 i f_1 + 2 a q &=& 0,\nonumber\\
g_{2x} - 2 i f_2 - 2 a r &=& 0, \label{N09}
\eqa
finally yielding the extended differential constraint:

\beq   \begin{array}{lc}
\hat{L}(g_1,g_2)+2i(r f_{1x}-q_x f_2)=0,\label{N10}
\end{array}   \eeq
with $\hat{L}(g_1,g_2)$ given in Eq. \ref{N05}.\\
\item The additional presence of ${\cal O}\left(\lambda^{-2}\right)$ terms in the zero-curvature condition yields a {\it second} constraint of the form,

\beq   \begin{array}{lc}
\hat{L} (f_1, f_2) = 0,
\end{array}   \eeq
with $f_{1,2}$ replacing $g_{1,2}$, respectively, in Eq. \ref{N05}.
\end{enumerate}
\paragraph*{}Thus, the perturbed NLS equations (Eqs. \ref{N03}) remain the same, although the order of the differential constraint is
increased recursively, thereby creating a new integrable hierarchy for the corresponding system. This is possible as the
NLS equations themselves are sensible exclusively to the ${\cal O}\left(\lambda^{-1}\right)$ contribution of $V_D(\lambda)$.
Any other {\it additional} deformations of ${\cal O}\left(\lambda^{-n}\right),~1<n\in\mathbb{Z}$ only constructs additional higher order constraints.

\paragraph*{} In the above a one-to-one correspondence is established between the NHDs of Lax pair and  Kupershmidt's formalism. In the prior the deformed equations and the constraint conditions both follow from a specific Lax pair which automatically points towards compatibility between the dynamical flows and the constraints imposed on them. Kupershmidt's method deals with the identification of such compatibility through several examples involving both continuous and discrete cases.

\paragraph*{}In the next section, the QID of the NLS system will be discussed, which preserves integrability
of the system only {\it partially}. A comparison of the same with the NH deformation will illustrate critical aspects of
integrability conditions of the concerned system.


\section{QID of the NLS equation}
Certain non-integrable models are known to possess physical properties similar to the integrable ones. This distinct class
is found to contain models that have solitonic solutions, with properties very similar to the integrable counterparts \cite{f1}. In 2+1 dimensions, such solitonic structures are observed in baby Skyrme model having many potentials
and in Ward-modified chiral model \cite{O2}. Therefore, recent attempts were made to model such systems as deformed version
of integrable ones, with partially conserved nature, called quasi-integrable (QI) systems \cite{1,f1}. They are viewed as
parametric generalizations of their integrable counterparts. Such a generalization manifests itself through the existence of the
functions $P_n$ \cite{O2} as,

$$\frac{dQ_n(t)}{dt}=P_n(t),\quad n\in {\mathbb Z},$$
wherein $Q_n$s are the `charges' that would have been conserved for the corresponding integrable system with vanishing
$P_n$s. In general, for quasi-integrable systems, only a subset of all $P_n$s vanish. However, {\it asymptotically},
all of them disappear rendering the systems integrable. In particular the two soliton
configuration has the property 

\beq
Q_n (t \to +\infty)-Q_n (t \to -\infty)=\int_{-\infty}^{\infty} dt\,P_n(t) = 0,
\eeq
corresponding to conserved asymptotic charges. For breather-like solutions, the corresponding asymptotic condition is
$Q_n (t) = Q_n (t + T )$. For other configurations, such as multi-soliton systems, $P_n$s do not vanish, but display
interesting boundary properties of topological nature \cite{O2}. Recently, the concept of quasi-integrability has also been
extended to suspersymmetric models \cite{Own}.

\paragraph*{}The quasi-deformed systems are usually obtained through deforming the `potential' (nonlinear) term of the concerned
integrable system perturbatively \cite{f1,O2}. However, this scheme of deformation may not be unique \cite{1}, especially
for the NLS system. Then the curvature function of Eq. \ref{ZC} is obtained to identify the anomaly function ${\cal X}$ such that $F_{tx}\propto{\cal X}\neq 0$ \cite{1}. In order to achieve this, gauge transformation is performed
under the characteristic $sl(2)$ loop algebra of the system and the equations of motion are utilized.

\subsection{The quasi-NLS Construction}
Our aim here is to obtain the analytic structure of the quasi-NLS equation, that can be compared with its NHD counterpart in
Eqs. \ref{N03}, \ref{N05} and \ref{N06}. Subjected to the liberty of choice in deformation of quasi-NLS system in Ref. \cite{1},
the deformation in the NLS potential:

\bqa
&&{\cal V}(q)\rightarrow {\cal V}(q,\varepsilon)=\frac{1}{2 + \varepsilon} \left( \vert q \vert^2 \right)^{2 + \varepsilon},\quad\varepsilon\in{\mathbb R},\label{1}\\
&&{\cal V}(q)\equiv {\cal V}(q,\varepsilon=0)=\frac{1}{2}\vert q\vert^4,\nonumber
\eqa
is adopted. Moreover, in the NHD case, only the temporal Lax component $V$ was deformed so that the reverse-scattering
properties remain unaffected. This essentially amounts to identifying the term(s) in $V$ responsible for the
non-linear term (potential) in the equation, which can naturally be interpreted as functional derivative(s) of the potential ${\cal V}$
with respect to the modulus of the NLS solution. Considering the defocussing case, the Lax pair for the quasi-NLS system can be expressed as \cite{1},

\bqa
&&U=i \lambda \sigma_3 +q \sigma_+ +q^* \sigma_-,\nonumber\\
&&V_Q=-i\left(\lambda^2+\frac{1}{2}\frac{\delta {\cal V}}{\delta\vert q\vert^2}\right)\sigma_3 + \lambda(q \sigma_ +q^*\sigma_-)\nonumber\\
&&\qquad\quad+ \frac{i}{2}\left(q_x\sigma_+ -q^*_x\sigma_-\right), \label{NN01}
\eqa
where we have taken the self-coupling strength to be unity. Indeed the derivative of the potential appears only in the temporal Lax component $V_Q$. The QID can be induced by substituting the deformed potential ${\cal V}(q,\varepsilon)$ in the expression.

\paragraph*{}Before going ahead with the formal comparison with NH deformation, it is fruitful to concisely review the
QID performed in Ref. \cite{1}. The anomalous charges were obtained through the standard Abelianization procedure
by performing gauge transformations based on the characteristic $sl(2)$ loop algebra of the NLS system, with the gauge-fixing
condition that the new temporal Lax component is defined in the Kernel subspace of the loop algebra. The quasi-conserved
charges $Q_n$ are found to satisfy,

\beq
\frac{dQ_n}{dt}=\int^\infty_{-\infty}dx\,{\cal X}\alpha_n,\label{N001}
\eeq
where ${\cal X}$ being the anomaly from the deformed curvature condition and $\alpha_n$s being the expansion coefficients of the same in the Kernel subspace,
{\it following} the gauge transformation. Then, utilizing the ${\mathbb Z}2$ symmetry ($sl(2)$ automorphism $\otimes$
space-time parity) of the system, it was shown that all $\alpha_n$s are parity even in time. Thus, for NLS, as ${\cal X}$ is parity-odd
by construction, the charges in Eq. \ref{N001} are conserved asymptotically (scattering limit) which ensures quasi-integrability
of the system. This result is general upto the potential being dependent on the modulus $\vert q\vert$, the latter being
parity-even, along with a phase that is parity-odd.

\paragraph*{}Following Eq. \ref{1} (or Eq. \ref{NN01}) the NLS equation gets quasi-deformed as,

\beq
q_t-\frac{i}{2}q_{xx}+i\vert q\vert^{2+2\varepsilon}q=0.\label{N002}
\eeq
The preceding discussion of NHD in Sec \ref{S2} suggests that the above system cannot be identified with QID absolutely.
By definition QID support a subset (infinite or finite) of charges which are anomalous whereas
NHD preserves {\it all} the charges of the initial system, albeit subjected to additional constraints. Equivalently, if the exact identification was possible then one should obtain $V_Q=\tilde{V}=V_O+V_D$. From Eq. \ref{NN01} that is possible iff,

\beq
\frac{\partial {\cal V}(q,\varepsilon)}{\partial\vert q\vert^2}\sigma_3=\vert q\vert^2\sigma_3+i2V_D.
\eeq
However, this contradicts with the prescription in Eq.s \ref{N102} as $V_D$ needs to be of ${\cal O}\left(\lambda^{-1}\right)$ and $g_{1,2}\neq 0$  to invoke NHD.
Even a wishful thinking like $\varepsilon=\varepsilon(\lambda)$ fails to compensate as one ends up with the requirement,

\beq
G^{(1)}=-\lambda\vert q\vert^2\sum_{n=1}^\infty\frac{\varepsilon^n}{n}\left(\log\vert q\vert^2\right)^n\sigma_3,\quad\varepsilon\to 0,\label{NN03}
\eeq
which was specifically required to be ${\cal O}\left(\lambda^0\right)$ for a meaningful NHD.
However, the condition of asymptotic integrability for the QID, especially
in case of NLS equation \cite{1}, strongly suggest a conditional equivalence. Therefore, it is logical to expect
certain limits to exist under which the QID-NHD correspondence can be realized.

\paragraph*{}We take the lead from the single soliton solution for the deformed potential in Eq. \ref{1} \cite{1}:

\bqa
&&q=\Big[(2+\varepsilon)\rho^2{\rm sech}^2\left\{(1+\varepsilon)\rho(x-vt-x_0)\right\}\Big]^{1/(1+\varepsilon)}\nonumber\\
&&\qquad\times\exp\Big\{i2\left(\rho^2t-\frac{v^2}{4}t+\frac{v}{2}x\right)\Big\},\label{N003}
\eqa
with $(\rho,v,x_0)\in{\mathbb R}$, which falls-back to the standard NLS bright soliton solution for $\varepsilon\to 0$.
Even otherwise, the `deformed' soliton has the same asymptotic behavior as the undeformed counterpart \cite{1}. We consider
the $\varepsilon\to 0$ approach to extract the NHD-QID correspondence {\it locally}. For comparison, the bright soliton
solution or a non-holonomic NLS system is given as \cite{Kundu01},

\bqa
&&q=2\rho_d^2{\rm sech}^2\left[\rho_d(x-v_dt-x_0)\right]\nonumber\\
&&\qquad\exp\left[i2\left(\rho_d^2t-\frac{v_d^2}{4}t+\frac{v_d}{2}x\right)\right],\label{N004}
\eqa
having velocity $v_d=v+v'$ with $v'=\tilde{c}(t)/t\vert\lambda_1\vert^2$ and frequency $\rho_dv_d=\rho v+\omega'$ with
$\omega'=-2\rho(x-vt)\tilde{c}(t)/t\vert\lambda_1\vert^2$. Here $\lambda_1=\rho(x-vt)+i\eta$, $\eta$ being a parameter of
deformation, and $\tilde{c}(t)$ is the asymptotic value of the perturbing function. It is clear that all these parameters
again approach their undeformed values asymptotically ($\vert x\vert\to\infty$). Therefore, it complements the parametric
limit $\varepsilon\to 0$ of QID asymptotically.

\paragraph*{}In the limit $\varepsilon\to 0$, the quasi-NLS system of Eq. \ref{N002} can be expanded as,

\bqa
&&q_t-\frac{i}{2}q_{xx}+i\vert q\vert^2q=-i\varepsilon\vert q\vert^2q\log(\vert q\vert^2)\nonumber\\
&&\qquad\qquad\qquad\qquad\quad-\frac{i}{2}\varepsilon^2\vert q\vert^2q\log^2(\vert q\vert^2)+{\cal O}(\varepsilon^3).\label{N005}
\eqa
The singularity of logarithms for $\vert q\vert^2\to 0$ is effectively regulated by the term $\vert q\vert^2q$ for first few values of $n$, the latter being the power of the logarithm ($\log^n(\vert q\vert^2)$). Beyond that a singularity is still assured to be avoided due to the factor $\varepsilon^n$. In fact, asymptotically speaking,
localized soliton solutions given in Eq.s \ref{N003} and \ref{N004} always attain infinitesimal but non-zero values in
physical systems, avoiding such singularities altogether. However, the QID in Eq. \ref{1} is {\it not} unique,
as the following choices of deformation,

\bqa
&&{\cal V}(q,\varepsilon)=\frac{1}{2}\vert q\vert^4+\varepsilon\vert q\vert^6\nonumber\\
{\rm or}\quad&& {\cal V}(q,\varepsilon)=\frac{1}{2}\vert q\vert^4\exp\left(-\varepsilon\vert q\vert^2\right),\label{N008}
\eqa
also works \cite{1}. Such deformations fall within the general premises of the Hamiltonian approach of QID \cite{Own2}. From the above deformations, one obtains the following quasi-NLS equations:

\bqa
&&q_t-\frac{i}{2}q_{xx}+i\vert q\vert^2q=-i\frac{3}{2}\varepsilon\vert q\vert^4q \quad{\rm and}\label{N009a}\\
&&q_t-\frac{i}{2}q_{xx}+i\vert q\vert^2q= i\frac{3}{2}\varepsilon\vert q\vert^4q-i\varepsilon^2\vert q\vert^6q+{\cal O}(\varepsilon^3),\label{N009b}
\eqa
respectively. It is to be noted that only for the second equation above, the limit $\varepsilon\to 0$ is necessary. In general the Eq.s \ref{N005}, \ref{N009a} and \ref{N009b} can be written as,

\bqa
&&q_t-\frac{i}{2}q_{xx}+i\vert q\vert^2q=-i\left(\frac{\delta{\cal V}(q,\varepsilon)}{\delta\vert q\vert^2}-\vert q\vert^2\right)q\nonumber\\
&&\qquad\qquad\qquad\qquad=q\int^x{\cal X}\quad{\rm where,}\label{41}\\
&&{\cal X}=-i\partial_x\left(\frac{\delta{\cal V}(q,\varepsilon)}{\delta\vert q\vert^2}-\vert q\vert^2\right),\nonumber
\eqa
is the QID anomaly mentioned above.

\paragraph*{}The last two examples reveal an important aspect. To obtain a quasi-NLS system in the form
$q_t-\frac{i}{2}q_{xx}+i\vert q\vert^2q\neq 0$, similar in form to the case of NHD of Eq.s \ref{N02}, it is not {\it always necessary}
to take the $\varepsilon\to 0$ limit. However, when $\varepsilon$ effects the {\it degree} of non-linearity, this limit
becomes important. It makes sense as a stable solution for a given non-linear system depends on the counterbalance
between dispersion and non-linearity, and when the order of the latter is changed the immediate stable structures that
{\it still} survive are small deviations of the original. This crucial aspect was highlighted in Ref. \cite{1} along with
numerical support. However, except for the need of a localized solution that has `sensible' asymptotic behavior, the
perturbative limit $\varepsilon\to 0$ is not necessary for quasi-deformation in general \cite{SSG0,f1,1,Own,Own2}. This appears quite
simply in case of the deformation of Eq. \ref{N009a}, wherein the $\varepsilon$ does not appear in the power of the
non-linear term and thus, the RHS is exact. Such a system with clearly {\it higher} order non-linearity can have stable
solutions, but are expected to be very different from the undeformed ones. However, in case of localized solution, the system can always be expected to match asymptotically with the undeformed one in the limit $\varepsilon\to 0$ owing to the smallness of $\vert q\vert$.

\section{Comparison between NHD and QID}
We have seen that NHD deforms the NLS system by introducing inhomogeneity yet preserves the integrability of the system through higher-order constraints among the deformation functions. On the other hand QID is implemented by modifying the inherent non-linearity in way that the system remains `partially integrable' in terms of remaining number of conserved charges. However, the fact that the latter deformation returns to complete integrability in the asymptotic limit renders the question whether these two classes of deformations have something in common. Albeit they cannot be absolutely identified as only NHD maintains absolute integrability.  
\paragraph*{}From the Lax pair in Eq. \ref{N01} the NHD can be achieved  both focusing ($r=-q^*$) and defocussing ($r=q^*$) NLS systems. However, the quasi-NLS systems were derived specifically for the {\it defocussing} case \cite{1} by modifying the self-interaction `potential' which only effected the modulus $\vert q\vert$ of the NLS solution but not its phase. For a comparison between NHD and QID of the NLS system we will consider the {\it defocussing} case explicitly from hereon. The focusing case corresponds to somewhat different QID treatment, following different asymptotic behavior for $x\to\pm\infty$ \cite{Focus}. 
\paragraph*{The phase discontinuity:}For $r=q*$ in case of NHDEq.s \ref{N03} and \ref{N04} leads to simpler results,

\beq
g_2=-g_1^*,\qquad a\in{\mathbb R}\quad{\rm and}\quad\frac{g_{1\,x}}{q}\in{\mathbb R},\label{N0010}
\eeq
not possible for the focusing analogue. The last of the above equations forces a constraint,

\bqa
&&\theta_x=\frac{R_x}{R}\tan(\phi-\theta);\quad g_1:=R\exp(i\theta)\nonumber\\
{\rm and }\quad&&q=\vert q\vert\exp(i\phi),\label{N0011}
\eqa
which in essence is equivalent to Eq.s \ref{N05} and \ref{N06}. Thus the non-holonomic constraint in this case
essentially becomes a {\it modulus-phase correlation} for the corresponding inhomogeneity parameter $g_1=-g_2^*$. The parameter $a$ is real and can be completely eliminated. Any function $g_1$ that satisfies Eq. \ref{N0011} is suitable for imposing NHD on the defocussing NLS system.
\paragraph*{}On eliminating the real parameter $a$ in Eq.s \ref{N06} for $r=q*$ and then separating real and imaginary parts eventually lead to a pair of equations:

\bqa
&&\frac{1}{\vert q\vert}\left(R_{xx}-(\theta_x)^2R\right)-\frac{1}{\vert q\vert^2}\left(R_x\vert q\vert_x-\theta_x\phi_x\vert q\vert R\right)\nonumber\\
&&\quad=-4R\vert q\vert\cos^2(\theta-\phi)\quad{\rm and}\nonumber\\
&&\frac{1}{\vert q\vert}\left(2\theta_xR_x+\theta_{xx}R\right)-\frac{1}{\vert q\vert^2}\left(\theta_xR\vert q\vert_x+R_x\phi_x\vert q\vert\right)\nonumber\\
&&\quad=2R\vert q\vert\sin 2(\theta-\phi).\label{NN06}
\eqa
These equations need to be satisfied simultaneously along with Eq. \ref{N0011}. This leaves enough room to choose $R$ as $R=R(\vert q\vert)$,
allowing for a possibility of identification with QID. One can hope that
the RHS of the a quasi-NLS system of the type in Eq. \ref{41} could be identified with $-g_1$. As the prior have the general form:
$$i\times({\rm modulus})\times\exp(i\phi)=({\rm modulus})\times\exp\left[i\left(\phi+\frac{\pi}{2}\right)\right],$$
such an identification would imply $\theta=\phi+\pi/2$. From Eq. \ref{N0011} this will mean that the phase of the deformation inhomogeneity is discontinuous as $\theta_x=\phi_x=\infty$. For such singular behavior, $R$ in Eq.s \ref{NN06} is now undetermined
and one can chose it to match QID. This highly non-trivial analytic condition for the phase of $g_1$ (and also that of the solution $q$) demonstrates the incompatibility between QID and NHD in general. Even $\theta_x=\infty$ the
`identification' of QID with NHD is superfluous as the quasi-integrability itself becomes undefined. This is because all the $\alpha_n$s in Eq. \ref{N001} for $n\ge3$ contains $\phi_x$ (Appendix A of Ref. \cite{1}) and become singular implying $d_tQ_n=\infty$. 
\paragraph*{}Although rare the observed overlap between non-holonomic and quasi-deformed NLS system mandates further clarification. This overlap can be straight-forwardly verified for the higher-order NHD induced by the deformations in Eq. \ref{N07}. 
The constraint conditions for NHD (Eq.s \ref{N04}), in principle, can be solved leaving out only one independent function among $a$, $g_1$ and $g_2$\footnote{For the defocussing case, those three conditions reduce to two as $g_2=-g_1^*$.}. Quantitatively, for 

\beq
g_1=i\left(\frac{\delta{\cal V}}{\delta\vert q\vert^2}-\vert q\vert^2\right)q,\label{NN10}
\eeq
the QID anomaly function can be expressed as,

\beq
{\cal X}=-\partial_x\left(\frac{g_1}{q}\right)\equiv-i\partial_x\left(\frac{R}{\vert q\vert}\right).\label{NN11}
\eeq
As ${\cal X}$ is parity-odd for parity-even $\vert q\vert$ (soliton case) \cite{1} we see that $R$ also needs to be parity-even for the NHD-QID correspondence. This conclusion is of considerable importance subjected to the asymptotic behavior as we will see next. Also, the QID parameter  $\varepsilon$ could be a local function (both in space and time) in general \cite{1} additionally requiring it to be parity even.
\paragraph*{The perturbative limit:}To illustrate the scenario more clearly, let us consider the perturbative limit of QID as it amounts to minimal deviation from integrability. In that case, an attempt to identify QID with NHD at the level of Lax pair, then from Eq. \ref{NN03} one gets, 

\beq
G^{(1)}\approx-\lambda\varepsilon\vert q\vert^2\log\left(\vert q\vert^2\right)\sigma_3,
\eeq
for $\varepsilon\to 0$. A simple redefinition $\varepsilon\to\epsilon/\lambda$ then implies,

\beq
a=-\varepsilon\vert q\vert^2\log\left(\vert q\vert^2\right),\quad g_{1,2}=0.\label{NN04}
\eeq
These results are not compatible with the general conditions in Eq.s \ref{N04} except for the trivial case $q=0$. 
\paragraph*{Gauge incompatibility:}QID manifests through modification of the Lax pair (Eq.s \ref{NN01}) exclusively in the Kernel subspace of the $SL(2)$ loop algebra which automatically implies $g_{1,2}=0$. On the other hand NHD maintains integrability through contribution from the Image subspace supported by additional constraints. The deformed dynamical equations in both cases, however, belong to the Image sector of the zero curvature condition.
This situation endorses the possibility that for $\theta=\phi+\frac{\pi}{2}$ the respective Lax components for QID and NHD are gauge-equivalent. In particular, this amounts to the conditions:

\bqa
&&G\tilde{V}G^{-1}+G_tG^{-1}=V_Q\nonumber\\
{\rm and}\quad&& GUG^{-1}+G_xG^{-1}=U,\label{NN12}
\eqa
to be satisfied.
The element of the $sl(2)$ gauge group can be parameterized as: 
\bqa
&&G=\exp\left(\boldsymbol{\alpha}\cdot\boldsymbol{\sigma}\right)\nonumber\\
{\rm with}\quad&&\boldsymbol{\alpha}\cdot\boldsymbol{\sigma}=\alpha_3\sigma_3+\alpha_+\sigma_++\alpha_-\sigma_-.\nonumber\\
{\rm Further},\quad &&G=\cosh\vert\boldsymbol{\alpha}\vert+\hat{\alpha}\cdot\boldsymbol{\sigma}\sinh\vert\boldsymbol{\alpha}\vert\nonumber\\
{\rm where}\quad&&\vert\boldsymbol{\alpha}\vert^2=\boldsymbol{\alpha}\cdot\boldsymbol{\alpha}.\label{NN13}
\eqa
Therefore, the gauge transformations take the explicit forms: 

\bqa
&&\hat{\alpha}\cdot\boldsymbol{\alpha}_x\,\hat{\alpha}\cdot\boldsymbol{\sigma}\nonumber\\
&&+\left[2\left(U-\hat{\alpha}\cdot\boldsymbol{U}\hat{\alpha}\right)+\frac{1}{\vert\boldsymbol{\alpha}\vert}\left(\hat{\alpha}\times\boldsymbol{\alpha}_x\right)\right]\cdot\boldsymbol{\sigma}\sinh^2\vert\boldsymbol{\alpha}\vert\nonumber\\
&&+\left[i\left(\hat{\alpha}\times\boldsymbol{U}\right)+\frac{1}{2\vert\boldsymbol{\alpha}\vert}\left(\boldsymbol{\alpha}_x-\hat{\alpha}\cdot\boldsymbol{\alpha}_x\,\hat{\alpha}\right)\right]\cdot\boldsymbol{\sigma}\sinh2\vert\boldsymbol{\alpha}\vert\nonumber\\
&&\qquad=0\nonumber\\
&&{\rm and}\nonumber\\
&&V_D+\hat{\alpha}\cdot\boldsymbol{\alpha}_t\,\hat{\alpha}\cdot\boldsymbol{\sigma}\nonumber\\
&&+\left[2\left(\tilde{V}-\hat{\alpha}\cdot\boldsymbol{\tilde{V}}\hat{\alpha}\right)+\frac{i}{\vert\boldsymbol{\alpha}\vert}\left(\hat{\alpha}\times\boldsymbol{\alpha}_t\right)\right]\cdot\boldsymbol{\sigma}\sinh^2\vert\boldsymbol{\alpha}\vert\nonumber\\
&&+\left[i\left(\hat{\alpha}\times\boldsymbol{\tilde{V}}\right)+\frac{1}{2\vert\boldsymbol{\alpha}\vert}\left(\boldsymbol{\alpha}_t-\hat{\alpha}\cdot\boldsymbol{\alpha}_t\,\hat{\alpha}\right)\right]\cdot\boldsymbol{\sigma}\sinh2\vert\boldsymbol{\alpha}\vert\nonumber\\
&&\qquad=\frac{1}{2}\int^x{\cal X}\sigma_3.\label{NN14}
\eqa
In the above $U=\boldsymbol{U}\cdot\boldsymbol{\sigma}$ and $\tilde{V}=\boldsymbol{\tilde{V}}\cdot\boldsymbol{\sigma},\quad\tilde{V}=V_0+V_D$ with $\hat{\alpha}=\boldsymbol{\alpha}/\vert\boldsymbol{\alpha}\vert$. Though the above equations looks complicated enough to solve for $\vert\boldsymbol{\alpha}\vert$, on isolating coefficients of $\sigma_{3,\pm}$ and then further isolating different orders of $\lambda$ yields a set of simpler equations. Particularly from the first equation at ${\cal O}\left(\lambda\right)$ the coefficients of $\sigma_3$ yields the result:

\beq
\sinh\vert\boldsymbol{\alpha}\vert=0,\quad{\rm implying}\quad\vert\boldsymbol{\alpha}\vert=in\pi,\quad n\in\mathbb{Z}.\label{NN15}
\eeq
This makes sense as being constructed out of $sl(2)$ gauge parameters $\alpha_{3,\pm}$, which are ought to be complex in general, the Euclidean product $\vert\boldsymbol{\alpha}\vert^2$ can very well be negative ({\it e. g.}, $-n^2\pi^2$). This greatly simplifies Eq.s \ref{NN14} to,

\beq
\hat{\alpha}\cdot\boldsymbol{\alpha}_x\,\hat{\alpha}\cdot\boldsymbol{\sigma}=0\quad{\rm and}\quad V_D+\hat{\alpha}\cdot\boldsymbol{\alpha}_t\,\hat{\alpha}\cdot\boldsymbol{\sigma}=\frac{1}{2}\int^x{\cal X}\sigma_3,\label{NN16}
\eeq
where we have utilized Eq. \ref{NN11}. The first equation essentially is the identity that $\vert\boldsymbol{\alpha}\vert_x=0$ following Eq. \ref{NN15}. Since $\vert\boldsymbol{\alpha}\vert$ is a constant, the coefficients of $\sigma_{3,\pm}$ in the second of Eq.s \ref{NN16} leads to the null result,

\beq
a=-i\lambda\int^x{\cal X},\quad g_{1,2}=0\quad\forall\alpha_{3,\pm},\label{NN17}
\eeq
which is exactly same as Eq. \ref{NN04}. So we conclude that even for the most general case ({\it i. e.}, arbitrary $\boldsymbol{\alpha}$ and non-perturbatively) QID and NHD cannot be gauge-equivalent\footnote{We have not considered the possibility $\boldsymbol{\alpha}=\boldsymbol{\alpha}(\lambda)$ here since gauge-invariance of the NLS and non-holonomic-NLS system does not permit it. However, for a more generalized NLS-like system this may be allowed which may disagree with the current null result but is highly unlikely.}. 
\paragraph*{Amplitude scaling:}The comparison of local deformations like QID of the NLS system with NHD can be extended to more arbitrary class of deformations in view of understanding their proximity to the NHD. As
an example, we consider the simple case of scaling the modulus of the NLS solution:

\bqa
&&\vert q \vert \rightarrow \Bigl[ 1 + f \left( \varepsilon, \vert q \vert \right) \Bigr] \vert q \vert\nonumber\\
{\rm and}\quad&&\vert q \vert^2 \rightarrow \Bigl[ 1 + g \left( \epsilon, \vert q \vert \right) \Bigr] \vert q \vert^2;\label{3}
\eqa
that complements the QID
of the defocussing NLS system with $\epsilon\to 0$, where,

\bqa
&&f\left( \epsilon, \vert q\vert \right) \approx \frac{1}{4} \Bigl( \varphi - \frac{1}{2} \Bigr)\epsilon + \frac{1}{32} \Bigl( \varphi^2-\varphi + \frac{5}{4} \Bigr)\epsilon^2\nonumber\\
&&\qquad\qquad+ {\cal O} \left( \epsilon^3 \right),\nonumber\\
&&g\left( \epsilon, \vert q \vert \right) \approx \frac{1}{2} \Bigl( \varphi - \frac{1}{2} \Bigr)\epsilon + \frac{1}{8} \Bigl( \varphi^2-\varphi + \frac{3}{4} \Bigr)\epsilon^2\nonumber\\
&&\qquad\qquad+ {\cal O} \left( \epsilon^3 \right);\label{4}\\
&&\varphi:=\log \left( \vert q \vert^2 \right).\nonumber
\eqa
Accordingly, the corresponding Lax pair components get deformed as,

\bqa
&&U \rightarrow \tilde{U}= -i \lambda \sigma_3 + \left(1 + f \right) (q \sigma_+ + q^* \sigma_-),\nonumber\\
&&V_O \rightarrow \tilde{V}= - i \lambda^2 \sigma_3 + \lambda \left(1 + f \right) (q \sigma_+ + q^* \sigma_-)\nonumber\\
&&\qquad\qquad\quad-\frac{i}{2}(1+g)\vert q \vert^2\sigma_3 \nonumber\\
&&\qquad\qquad\quad+ \frac{i}{2}(1+f)(q_x \sigma_+-q^*_x \sigma_-)\nonumber\\
&&\qquad\qquad\quad+\frac{i}{2} f_x (q \sigma_+ - q^* \sigma_-).\label{9}
\eqa
This further effects the scattering properties of the original system, unlike in NHD, as the spatial
Lax component is also deformed. As we will see in the following, this essentially means the correspondence to NHD in this case will also be approximate.
On imposing the zero curvature condition, the independent equations turn out to be,

\bqa
&& q_t - \frac{i}{2} q_{xx} + i (1 + g) \vert q \vert^2 q \nonumber\\
&&\qquad=(1 + f)^{-1}\left[ i f_x q_x + \frac{i}{2} f_{xx} q -f_t q\right], \nonumber\\
&& q^*_t + \frac{i}{2} q^*_{xx} - i (1 + g) \vert q \vert^2 q^* \nonumber\\
&&\qquad= -(1 + f)^{-1}\left[i f_x q^*_x + \frac{i}{2} f_{xx} q^* +f_t q^*\right]\nonumber\\
{\rm and}&&\nonumber\\
&& \Big [ (1 + f) \vert q \vert^2 \Big ]_x + f_x \vert q \vert^2 \nonumber\\
&&\qquad= (1 + f)^{-1}\Big [ (1 + g) \vert q \vert^2 \Big ]_x. \label{15}
\eqa
The last equation above reduces to a mere identity for $f,~g=0$, and the first two equations reduces to the usual defocussing NLS
equation and its complex conjugate. As $f$ and $g$ are small (Eq. \ref{4}), the
explicit form of the deformed equations are:

\bqa
&&q_t - \frac{i}{2} q_{xx} + i \vert q \vert^2 q \approx \epsilon{\cal A}+\epsilon^2{\cal B};\quad{\rm where},\nonumber\\
&& {\cal A}=\Big [ \frac{1}{4 \vert q \vert^2} \Big ( i \vert q \vert^2_x q_x + \frac{i}{2}\vert q \vert^2_{xx} q-\vert q \vert^2_t q\Big ) \nonumber\\
&&\qquad-\frac{i}{8\vert q \vert^4} \left ( \vert q \vert^2_x\right )^2 q - \frac{i}{2} \Big ( \log \left ( \vert q \vert^2 \right ) - \frac{1}{2} \Big ) \vert q \vert^2 q \Big], \nonumber\\
&& {\cal B}=\frac{1}{8} \Big [\frac{i}{2\vert q \vert^4} \Big ( \log \left ( \vert q \vert^2 \right ) - 1\Big ) \left ( \vert q \vert^2_x\right )^2 q \nonumber\\
&&\qquad- \Big ( \log^2 \left ( \vert q \vert^2 \right ) - \log \left ( \vert q \vert^2 \right ) + \frac{3}{4}\Big ) \vert q \vert^2 q \Big ], \label{16}
\eqa
and its complex conjugate. It is clear that if the deformation of Eq. \ref{3} is kept only at the level of the potential (Eq. \ref{1}), then only
the last terms in the square brackets on the RHSs in Eqs. \ref{16} would have survived, which is the case for QID.
\paragraph*{}Though ${\cal A}$ and ${\cal B}$ contain singular functions like logarithms of the NLS modulus $\vert q\vert$,
they are regulated by suitable powers of $\vert q\vert$ in the denominators. Therefore, the expansion is indeed perturbative.
This additionally ensures that the system will approach integrability asymptotically, as  ${\vert q\vert}$ is physically
expected to be well-behaved there.
\begin{figure*}
\centering
\begin{subfigure}[b]{0.58\textwidth}
\centering
\includegraphics[width=\textwidth]{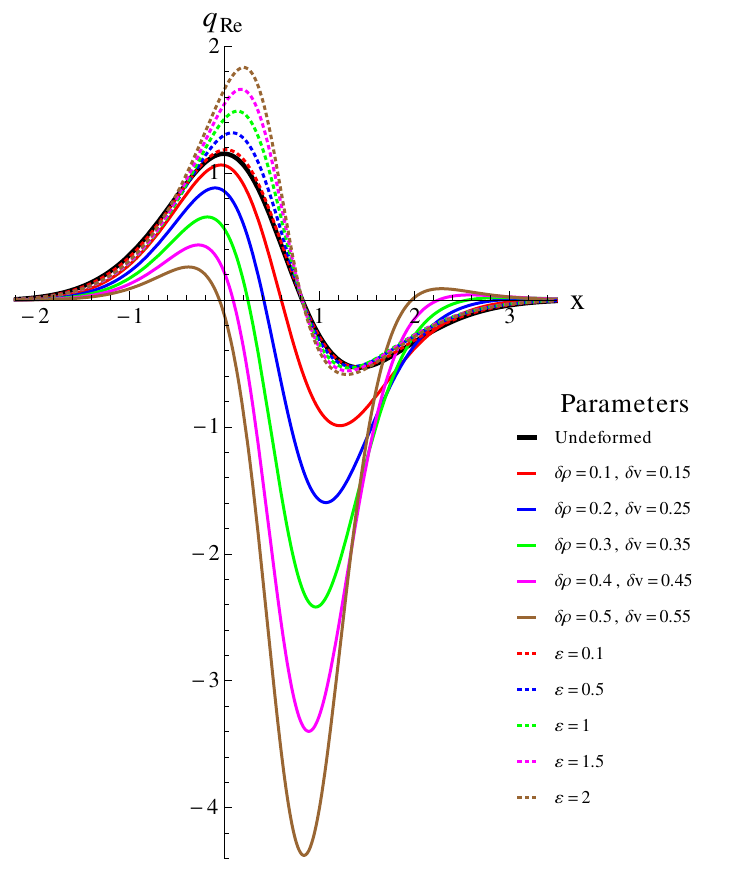}
\caption{Real part.}
\label{NRX}
\end{subfigure}
\begin{subfigure}[b]{0.4\textwidth}
\centering
\includegraphics[width=\textwidth]{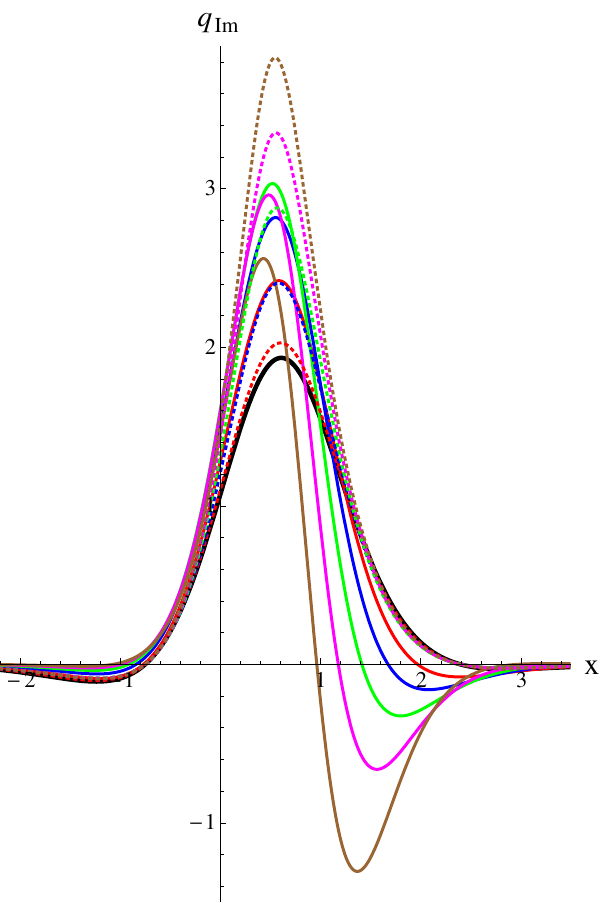}
\caption{Imaginary part.}
\label{QRX}
\end{subfigure}
\caption{The real and imaginary parts of the localized solutions for both NHD (Eq. \ref{N003}, shown in solid lines) and QID (Eq. \ref{N004}, shown in dotted lines) of NLS system as functions of space ($x$). We have set $\rho=1=v$ and $x_0=0$ with $t=0.5$ for the above plots. As the respective deformation parameters for NHD ($\delta\rho$ and $\delta v$) and QID ($\varepsilon$) increase (from red to brown) the local parts of the solutions deviate more and more from the undeformed solution (solid black line) in `opposite' directions and thus can never be identified with each-other. However for larger $x$ they tend to reach a common plateau and can be expected to coincide for $x\to\pm\infty$.}
\label{F1}
\end{figure*}
\paragraph*{}Evidently, the RHS of Eq. \ref{16} does not have the same phase $\phi+\frac{\pi}{2}$ as for QID and therefore can satisfy
the NHD constraint for $\theta-\phi\neq\frac{\pi}{2}$ and thus can non-trivially be identified as modulus $R$ of $g_1$. Therefore, similar to the quasi-NLS case it is tempting to identify this local scaling as a conditional NHD though the spatial Lax component $\tilde{U}$ is also changed. For only $\vert q\vert^2$ being scaled but not $\vert q\vert$, {\it i. e.} $f(\epsilon,\vert q\vert)=0,~g(\epsilon,\vert q\vert)\neq 0$, this scaling can take the form of a QID as only the Kernel sector of the temporal Lax component will be effected.

\subsection{The Asymptotic Behavior}
If $g_1$ is localized and asymptotically well-behaved,
then for $(x,t)\to\pm\infty$, $R$ assumes fixed value(s) and thereby $R_x,\,R_t\to 0$. This condition particularly effects
Eq. \ref{N0011} as now $\theta_x=0$ identically. Still the phase $\phi$ of the deformed solution cannot be space-dependent as for $R_x=0,\,\theta_x=0$ Eq.s \ref{NN06} implies,
\beq
\theta=\phi+(2n+1)\frac{\pi}{2},\quad n\in\mathbb{Z}.\label{NN09}
\eeq
Hence the non-holonomic NLS solution attains a constant phase at the spatial infinity and such a solution can now be identified with the integrable limit of a quasi-NLS system. Additionally, as now $\theta=\phi+\frac{\pi}{2}$ modulo a multiple of $2\pi$ for $n$ being even, the quasi-NLS case stated before can be identified with the NHD. For a localized solution $\vert q\vert_x(x\to\pm\infty)=0$ the QID anomaly in Eq. \ref{41} vanishes and the charges $Q_n$s are trivially conserved. Therefore the $\varepsilon$-dependent part of the quasi-NLS equation can now be identified with the NHD inhomogeneity $g_1$ safely. As for the scaling deformation, the RHS of Eq. \ref{16} differs from $\phi$ in a more complicated way than that for the QID case. Still it can be assumed that the deformation parameter $\epsilon$ can take complex values so that there is an overall phase on the RHS of the form $\phi+(2n+1)\frac{\pi}{2}$ making the identification with NHD possible.
\begin{figure*}
\centering
\begin{subfigure}[b]{0.58\textwidth}
\centering
\includegraphics[width=\textwidth]{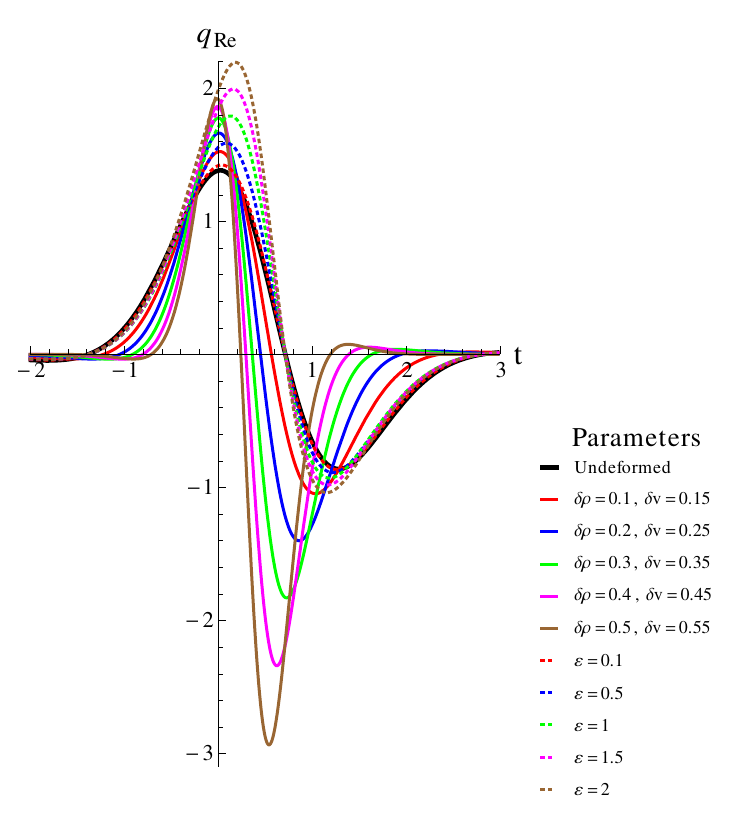}
\caption{Real part.}
\label{NRT}
\end{subfigure}
\begin{subfigure}[b]{0.4\textwidth}
\centering
\includegraphics[width=\textwidth]{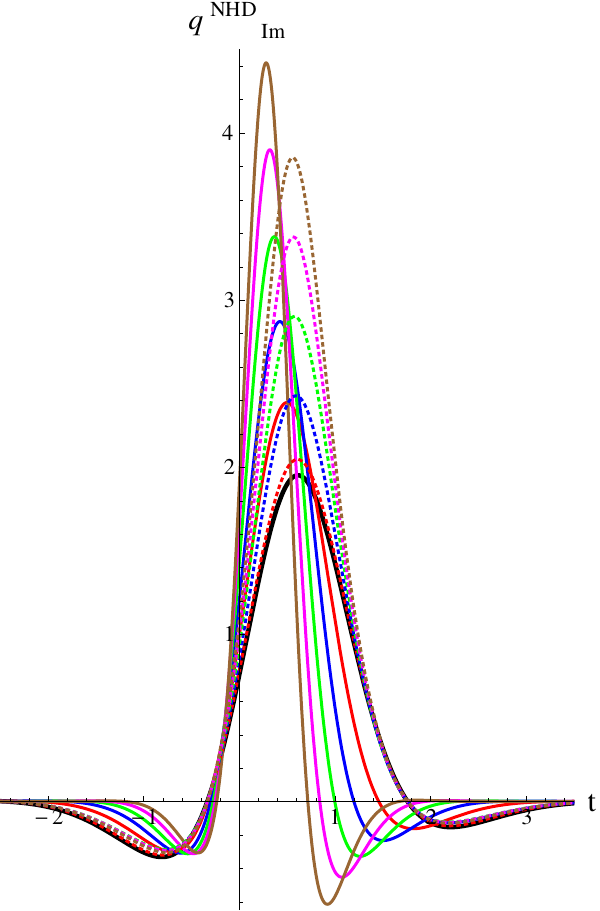}
\caption{Imaginary part.}
\label{QRT}
\end{subfigure}
\caption{The real and imaginary parts of the localized solutions for both NHD (Eq. \ref{N003}, shown in solid lines) and QID (Eq. \ref{N004}, shown in dotted lines) are varied with time. Here we have considered $\rho=1=v$ and $x_0=0$ with $x=0.5$. Interestingly with the increase in deformation parameters (from red to brown) the local deviations due to NHD and QID do not move away from the undeformed solution (solid black line) in the opposite direction unlike the $x$-variation, rendering possible convergence for certain parameter values including that of $x$. Asymptotically, however, ($t\to\pm\infty$) both classes of solution tend to converge.
}
\label{F2}
\end{figure*}
\paragraph*{}The above inferences are in conformity to the localized solutions for quasi-deformed and non-holonomic NLS system, given in Eq.s \ref{N003} and \ref{N004} respectively, which have have been plotted for a clearer picture.
In Fig. \ref{F1} the real parts of the localized (solitonic) solutions for both the cases are plotted with respect to space coordinate. Their local difference is considerable which becomes more prominent as the deviation parameters ($\rho_d-\rho$, $v_d-v$ for NHD and $\varepsilon$ for QID) increase depicting mutual incompatibility. However, they tend to converge more and more as $x\to\pm\infty$. The same behavior is seen for the respective imaginary parts in Fig \ref{F2}.
\paragraph*{}The case of temporal asymptot ($t\to\infty$) is not as directly apparent as the spatial counterpart since NHD conditions do not contain time derivatives by construction. Still we may assume $R_t(t\to\infty)=0$ for a $g_1$ that is also temporally localized. A further assumption of $\theta(x,\,t\to\infty)=\theta(t)$ and $\phi(x,\,t\to\infty)=\phi(t)$ for a system that deviates a little from a non-dissipative (integrable) system. This does not conflict with the restricted identifications we have encountered of QID and scaling deformation. Particularly for QID as integrability is regained for $t\to\pm\infty$ the system is likely to be of non-holonomic NLS type at that limit in general. The NHD formalism is tailored not to effect the time-evolution of the particular system and thus a non-holonomic system should not change at the temporal infinity. In that case as ${\cal X}$ is restricted to the Kernel subspace of the algebra and as it does {\it not} vanish at $t\to\pm\infty$ in general it is viable to identify ${\cal X}(x,\,t\to\pm\infty)$ as a non-holonomic deformation. This in effect translates to a complementing behavior of the deformation parameter $\varepsilon$ at the temporal infinity. 
\paragraph*{}The quasi and non-holonomically deformed soliton solutions of Eq.s \ref{N003} and \ref{N004} respectively are now plotted as a function of time in Fig. \ref{F2}. Though both the solutions deform more and more with the increase in the deformation parameters, their {\it relative} deviation does not increase in tandem unlike the $x$-variation. This might be attributed to the fact that NHD does not effect the time-evolution of the system whereas QID only
effects the modulus $\vert q\vert$ of the solution. Therefore for certain fixed value of $x$, with suitable parameterization the time-profile of the respective solutions may coincide even locally\footnote{The exceptional condition $\theta_x=\infty$ for NHD-QID overlap could demonstratively incorporated given $x$ is fixed.}. Asymptotically both the solutions tend to converge as inferred.
\paragraph*{}In light of the preceding discussion the deviation from integrability is compensated by suitable contribution from both Image and Kernel subspaces in case of NHD. In case of QID and scaling deformation there is no such compensation, the latter even effecting the scattering data through deforming $U$. What happens asymptotically that all these three deformations converges to a simpler case. In case for NHD the Kernel contribution $a$ vanishes as the Image one ($g_1$) is fixed. This leaves a simple phase relation to be satisfied by QID and scaling deformation to get identified as NHD.
Further, the non-holonomic constraints lives at a {\it different} spectral order than the equation(s) of motion. In comparison, the QID anomaly ${\cal X}$ lives in the {\it same} spectral order as the scaling deformation. This supports the intuition that the deformation parameter for the latter two should be $\varepsilon=\varepsilon(\lambda)$ and $\epsilon=\epsilon(\lambda)$ respectively for identification with NHD. However, this requires an asymptotic limit as only then the Kernal subspace contribution to NHD becomes trivial.
\paragraph*{}Therefore, we see that deformations (QID and the particular local scaling of the modulus) of the defocussing NLS system can {\it conditionally} be identified with the corresponding NHD. QID can locally be identified for the {\it exceptional} case of $\theta_x=\infty$\footnote{Which might be in terms of the time-profile.} and asymptotically ($x,\,t\to\pm\infty$) with a more relaxed condition (Eq. \ref{NN09}) as the required constraints becomes trivial. This is independent of whether the QID parameter $\varepsilon$ is perturbative or not. On the other hand the scaling deformation needs to be perturbative to be identified even asymptotically as otherwise it may even have non-local features (Eq.s \ref{15}).

\section{Conclusion}
We have demonstrated, through explicit analysis for both NHD and QID of the defocussing NLS system that these two classes
have distinct analytic structures. The NHD induces inhomogeneity in the dynamical equation in addition to higher order
differential constraints at a different spectral order, thereby preserving the integrability. We have shown the agreement between the bi-Hamiltonian and the Lax pair methods of inducing NHD to the NLS system.
The QID disrupts integrability {\it locally} by making a subset of the charges anomalous. It can return to complete integrability as
a NHD when the phases of NLS solution and non-holonomic inhomogeneity satisfy at exceptional condition of non-local phase which shows them to be not gauge-equivalent. However, both these deformations can asymptotically be identified given the locality of the NLS solution and the NHD inhomogeneity. This was intuitively expected as the quasi-deformed systems are already known to regain integrability in the asymptotic limit. It will be of interest to extend this approach to other integrable systems, with additional
focus on more general (multi-soliton) solution domains and higher-order hierarchies.

\section*{Acknowledgement}
The authors are grateful to Professors Luiz. A. Ferreira, Wojtek J. Zakrzewski and Betti Hartmann  for their encouragement,
various useful discussions and critical reading of the draft. PG's research is partially supported by FAPESP through grant numbered 2016/06560-6. KA's research is done at and supported by the IF, Naresuan University, Thailand.


\begin{thebibliography}{}
%
\bibitem{Das}A. Das, {\em Integrable models} (World Scientific, Singapore 1989).
\bibitem{Z1}P. D. Lax, {\em Integrals Of Nonlinear Equations Of Evolution And Solitary Waves}, Commun. Pure Appl. Math. {\bf 21}, (1968) 467.
\bibitem{Z2}V. E. Zakharov and A. B. Shabat, {\em Exact theory of two-dimensional self-focusing and one-dimensional self-modulation of waves in nonlinear media},
Zh. Exp. Teor. Fiz. {\bf 61}, (1971) 118 [Soviet Phys. JETP {\bf 34}, (1972) 62].
\bibitem{SSG0}S. Ferrara, L. Girardello and S. Sciuto, {\em An infinite set of conservation laws of the supersymmetric sine-Gordon theory}, Phys. Lett. B {\bf 76}, (1978) 303.
\bibitem{f1}L. A. Ferreira and W. J. Zakrzewski, {\em The concept of quasi-integrability: a concrete example}, JHEP {\bf 2011}, (2011) 130.
\bibitem{New1}H. Blas and M. Zambrano, {\em Quasi-integrability in the modified defocusing non-linear Schr\"odinger model and dark solitons}, JHEP {\bf 2016}, (2016) 005.
\bibitem{New01}H. Blas and H. F. Callisaya, {\em Quasi-integrability in deformed sine-Gordon models and infinite towers of conserved charges}, Commun. Nonlinear
Sci. Numer. Simulat. {\bf 55}, (2018) 105.
\bibitem{New2}H. Blas, A.C.R. do Bonfim and A.M. Vilela, {\em Quasi-integrable non-linear Schrödinger models, infinite towers of exactly conserved charges and bright solitons}, JHEP {\bf 2017}, (2017) 106.
\bibitem{1}L. A. Ferreira, G. Luchini and W. J. Zakrzewski, {\em The concept of quasi-integrability for modified non-linear Schrödinger models}, JHEP {\bf 2012}, (2012) 103.
\bibitem{KK}  A. Karasu-Kalkani, A.Karasu, A. Sakovich, S.Sakovich and R.Turhan {\em  A new integrable generalization of the Korteweg-de Vries equation.}, J. Math. Phys. {\bf 49}, (2008) 073516.
\bibitem{Kuppershmidt}B. A. Kupershmidt, {\em KdV6: An integrable system}, Phys. Lett. A {\bf 372}, (2008) 2634.
\bibitem{3} A. Kundu, {\em Two-fold integrable hierarchy of nonholonomic deformation of the derivative nonlinear Schr\"odinger and the Lenells–Fokas equation}, J. Math. Phys. {\bf 51}, (2010) 022901.
\bibitem{GuhaJPA} P. Guha, {\em Nonholonomic deformation of generalized KdV-type equations}, J. Phys. A: Math. Theor. {\bf 42}, (2009) 345201.
\bibitem{GuhaRMP} P. Guha, {\em Nonholonomic deformation of coupled and supersymmetric KdV equations and Euler–Poincar\'e–Suslov method}, Rev. Math. Phys. {\bf 27}, (2015) 1550011.
\bibitem{FW}A. L. Fetter and J. D. Walecka, {\em Quantum Theory of Many-Particle System}, (Dover Publications, New York 2012).
\bibitem{Laxmanan}M. Lakshmanan, {\em Continuum spin system as an exactly solvable dynamical system}, Phys. Lett. A {\bf 61}, (1977) 53.
\bibitem{Hasimoto}R. Hasimoto, {\em A soliton on a vortex filament}, J. Fluid Mechanics {\bf 51}, (1972) 477.
\bibitem{Balakrishnan1}R. Balakrishnan, {\em On the inhomogeneous Heisenberg chain}, J. Phys. C: Solid State Phys.{\bf 15}, L1305 (1982); R. Balakrishnan, {\em Dynamics of a generalised classical Heisenberg chain}, Phys. Lett. {\bf 92}, (1982) 243.
\bibitem{Balakrishnan2}R. Balakrishnan, {\em Inverse spectral transform analysis of a nonlinear Schr\"odinger equation with x-dependent coefficients}, Physica D {\bf 16}, (1985) 405.
\bibitem{OwnEPJB}K. Abhinav and P. Guha, {\em Inhomogeneous Heisenberg spin chain and quantum vortex filament as non-holonomically deformed NLS systems}, Eur. Phys. J. B {\bf 91}, (2018) 52.
\bibitem{Shiva}B.K. Shivamoggi, {\em Vortex motion in superfluid $^4$He: effects of normal fluid flow}, Eur. Phys. J. B {\bf 86}, (2013) 275; R. A. Van Gorder, {\em Quantum Hasimoto transformation and nonlinear waves on a superfluid vortex filament under the quantum local induction approximation}, Phys. Rev. E {\bf 91}, (2015) 053201.
\bibitem{XXZKdV}J. Nian, {\em Note on Nonlinear Schrödinger Equation, KdV Equation and 2D Topological Yang-Mills-Higgs Theory}, arXiv:1611.04562[hep-th] (2016).
\bibitem{NHD1}O. Krupkov\'a, {\em Mechanical systems with nonholonomic constraints}, J. Math. Phys. {\bf 38}, (1997) 5098.
\bibitem{GS} Jean-Loup Gervais and Mikhail V Saveliev {\em Higher grading generalizations of the Toda systems}, Nucl. Phys. B {\bf 453}, Issues 1-2, (1995) 449.
\bibitem{FGGS} Luiz A. Ferreira, Jean-Loup Gervais, Joaquin Sanchez Guillen and Mikhail V Saveliev {\em Affine Toda systems coupled to matter fields}, Nucl. Phys. B {\bf 470}, Issues 1-2, (1995) 236.
\bibitem{Kundu01}A. Kundu, {\em Nonlinearizing linear equations to integrable systems including new hierarchies with nonholonomic deformations}, J. Math. Phyis. {\bf 50}, (2009) 102702.
\bibitem{Fordy}A. P. Fordy and D. D. Holm, {\em A tri-Hamiltonian formulation of the self-induced transparency equations}, Phys. Lett. A {\bf 160}, (1991) 143.
\bibitem{Kundu02}A. Kundu, {\em  Non-holonomic deformation of the DNLS equation for controlling
optical soliton in doped fibre media
}, IMA Journal of Applied Mathematics {\bf 77}, (2012) 382.
\bibitem{NKKS} M Nakazawa, Y Kimura, K Kurokawa, K Suzuki {\em Self-induced-transparency solitons in an erbium-doped fiber waveguide}, Physical Review A, {\bf 45(1)}, (1992) 23.
\bibitem{NYK} M. Nakazawa, E. Yamada, and H. Kubota {\em Coexistence of self-induced transparency soliton and nonlinear
Schrödinger soliton}, Phys. Rev. Lett. {\bf 66}, (1991) 2625.
\bibitem{O2}L. A. Ferreira, G. Luchini G. and W. J. Zakrzewski, {\em The Concept of Quasi-Integrability}, Nonlinear and Modern Mathematical Physics, AIP Cof. Proc., {\bf 1562}, (2013) 43.
\bibitem{Own}K. Abhinav and P. Guha, {\em Quasi-Integrability in Supersymmetric Sine-Gordon Models}, EPL {\bf 116}, (2016) 10004.
\bibitem{Own2}K. Abhinav and P. Guha, {\em Quasi-Integrability of The KdV System}, arXiv:1612.07499[math-ph].
\bibitem{Focus}C. Sulem and P.-L. Sulem, {\em The nonlinear Schr\"odinger equation: self-focusing and wave collapse}, (Springer-Verlag, Berlin 1999).



\end{thebibliography}
\end{document}